\documentclass[letterpaper, 10 pt, conference]{ieeeconf}  

\IEEEoverridecommandlockouts                              

\usepackage{graphicx} 
\usepackage[caption=false,font=footnotesize]{subfig}
\usepackage{epsfig} 
\usepackage[dvipsnames]{xcolor}
\usepackage{times} 
\usepackage{amsmath} 

\usepackage{amsthm}

\usepackage{amssymb}  
\usepackage{bbm}
\usepackage{comment}
\usepackage{mathtools}
\usepackage{algorithm}
\usepackage{algpseudocode}

\usepackage{soul}

\usepackage[toc,sort=use]{glossaries}
\usepackage{array}
\newglossary[slg]{symbols}{sym}{sbl}{List of Symbols}
\newglossary[slg]{hidden}{sym}{sbl}{Hidden Symbols}
\makeglossaries
\newcommand{\glom}[4]{
	\newglossaryentry{#1}
	{
	  name={\ensuremath{#2}},
	  symbol={\ensuremath{#3}},
	  description=#4,
	  type=symbols,
	}
}

\glom{error}{e}{e}{Error}

\glom{cap}{\mathbf{Q}}{\mathbf{Q}}{Capability Matrix}
\glom{cap_bin}{\mathbf{\overline Q}}{\mathbf{\overline Q}}{Binary Capability Matrix}
\glom{task_req}{\mathbf{Y}_m^*}{\mathbf{Y}_m^*}{Task Requirement Matrix index m}

\newcommand{\sy}[1]{\glssymbol{#1}}

\newtheorem{definition}{Definition}

\newtheorem{example}{Example}

\newcommand{\mc}[1]{\mathcal{#1}}
\newcommand{\mb}[1]{\mathbf{#1}}
\newcommand{\ch}[1]{\textcolor{black}{#1}}
 \newcommand{\todo}[1]{}
\DeclareMathOperator*{\minimize}{minimize~}
\DeclareMathOperator*{\subjto}{subject\,to~}
\title{\LARGE \bf
Resilient Task Allocation in Heterogeneous Multi-Robot Systems}

\author{Siddharth Mayya$^{1}$, Diego S. D’antonio$^{2}$, David Salda\~{n}a$^{2}$, Vijay Kumar$^{1}$ 
\thanks{This research was sponsored by the Army Research Lab through ARL DCIST CRA W911NF-17-2-0181.}
\thanks{$^{1}$S. Mayya and V. Kumar are with the GRASP Laboratory, University of Pennsylvania, Philadelphia, PA, USA {\tt\small\{mayya, kumar\}@seas.upenn.edu.}}%
\thanks{$^{2}$D. D'antonio and D. Salda\~{n}a are with the Autonomous and Intelligent Robotics Lab (AIRLab) at Lehigh University, Bethlehem, PA, USA {\tt\small saldana@lehigh.edu.}}%
}

\begin{document}

\maketitle
\thispagestyle{empty}
\pagestyle{empty}

\begin{abstract}

For a multi-robot system equipped with heterogeneous capabilities, this paper presents a mechanism to allocate robots to tasks in a resilient manner when anomalous environmental conditions such as weather events or adversarial attacks affect the performance of robots within the tasks. Our primary objective is to ensure that each task is assigned the requisite level of resources, measured as the aggregated capabilities of the robots allocated to the task. By keeping track of task performance deviations under external perturbations, our framework quantifies the extent to which robot capabilities (e.g., visual sensing or aerial mobility) are affected by environmental conditions. This enables an optimization-based framework to flexibly reallocate robots to tasks based on the most degraded capabilities within each task. In the face of resource limitations and adverse environmental conditions, our algorithm minimally relaxes the resource constraints corresponding to some tasks, thus exhibiting a graceful degradation of performance. Simulated experiments in a multi-robot coverage and target tracking scenario demonstrate the efficacy of the proposed approach.  

\end{abstract}

\section{INTRODUCTION} \label{sec:intro}

In recent years, heterogeneous multi-robot systems have demonstrated a potential to achieve complex real-world objectives due to their versatility in accomplishing specialized tasks which might require collaboration among different types of robots, e.g.~\cite{iocchi2003distributed,gunn2015dynamic,rizk2019cooperative}.
A crucial step towards achieving such behaviors is multi-robot task allocation (MRTA), which concerns itself with allocating robots to tasks in such a way that the resources required to execute the tasks successfully are made available (see~\cite{taxonomy,taxonomy2,khamis2015multi} for a taxonomy and survey of the topic). For instance, a possible approach is to classify the robots according to their heterogeneous capabilities (e.g., speed, sensor range, battery life, etc), and then assign aggregated capabilities to each task, based on given specifications~\cite{prorok2017impact,ravichandar2020strata}. \par 

For heterogeneous multi-robot systems operating in dynamic and complex environments, the diversity in the capabilities of the robots presents another advantage---\emph{resilience}: the ability to continuously operate and recover from failures with limited resources, e.g.~\cite{ramachandran2019resilience,saulnier2017resilient}. In our context, when a multi-robot system experiences difficulties in executing tasks due to changing environmental conditions or certain types of adversarial attacks, reallocating robots to tasks can significantly improve their performance as a team, e.g.,~\cite{emam2020adaptive}. Such a reallocation can take different forms, based on the type of failure that has occurred. For instance, if a team of ground robots tasked with surveilling an area encounters slippery terrain, a reallocation of aerial robots to the task might be desirable. However, if an adversarial attack were to reduce the effective communication range of the ground robots, supplying additional robots of the same kind to act as intermediate communication links might be a better solution. 
Note that, in these scenarios, specific capabilities of the robots were affected by disturbances, i.e., ground mobility and communication range, respectively. Hence, a way to facilitate effective and resilient task allocation is by: \textit{i)} identifying the extent to which the \emph{robot capabilities} within each task are affected; and  \textit{ii)} performing a suitable reallocation which ensures progress in each of the tasks. \par 

In this paper, we propose a novel heterogeneous multi-robot task allocation framework which explicitly quantifies the extent to which robot capabilities---pertaining to relevant aspects of the robots' operation such as ground speed or sensor coverage---are degraded by environmental disturbances. The primary objective of our optimization-based formulation is to allocate a team of robots to a set of given tasks in a deterministic manner such that constraints on the minimum aggregate capability requirements for each task are satisfied. Distinct from previous works in the literature~\cite{prorok2017impact,ravichandar2020strata,notomista2019optimal}, we impart resilience to our framework in two ways. First, we explicitly model the fact that, a given task can be accomplished via multiple possible combinations of robot capabilities---one of which can be selected based on the extent to which robot capabilities have been degraded by environmental disturbances. This achieves \emph{resilience via reconfiguration}---by allowing the algorithm to move robots to tasks where they can contribute the most. Second, in situations where the capabilities of the robots are too degraded to satisfy the requirements for all the tasks, we allow the algorithm to relax the capability requirement constraints for some tasks, to ensure that constraints corresponding to higher priority tasks continue to be met. Such a \emph{graceful degradation of performance} ensures that infeasible task allocation specifications in the face of significant environmental disturbances are handled effectively. \par

Leveraging robot heterogeneity in MRTA problems has classically been approached by scoring the ability of each robot to perform different tasks~\cite{notomista2019optimal,parker1994heterogeneous,marcolino2013multi}, and by explicitly enumerating the various task-related capabilities of the robots~\cite{prorok2017impact,vig2006multi}. The above discussed features of the proposed task allocation algorithm are owed to a quantifiable understanding of how different robot capabilities are degraded due to changing environmental conditions. \ch{In this paper, the current state of the multi-robot task is encoded via a scalar \emph{task-value function} which takes as inputs the states of the robots involved in the task, e.g.,~\cite{cortes2004coverage,pimenta2009simultaneous,oh2015survey}. We assume that environmental disturbances and adversarial attacks manifest themselves as unmodeled disturbances in the dynamics of the robots, which might affect the task-value function as well. At every point in time, we allow each robot to measure the discrepancy between the expected and measured progress that it makes towards modifying the task-value function.} A similar approach is presented in~\cite{emam2020adaptive}, where the real-time performance of robots at tasks is used to modify the suitability of robots towards different tasks. In this paper,  we instead leverage the heterogeneity model to identify which \emph{capabilities} are primarily responsible for the observed performance deviations---thus allowing the algorithm to make more expressive reallocation decisions. \par

The capability degradation metrics are then leveraged by a centralized mixed-integer quadratic program (MIQP) which \textit{i)} selects a capability configuration that is best suited for each task, \textit{ii)} generates the robot-to-task allocations to meet the requirements set by the chosen configuration, and \textit{iii)} violates the resource constraints for some tasks \ch{to the least extent possible (in a Pareto-optimal sense)}, if required. Our framework deploys robots in a resource-aware manner, by minimizing the team size \ch{(in a Pareto-optimal sense)} and allowing the mission designer to specify a cost of deployment for each type of robot. Similarly, the mission designer can also specify which tasks are less critical to the mission than others (and hence should be degraded in quality first). Lastly, robots experiencing high performance degradation---based on a user defined threshold---are automatically excluded from the allocation process.   \par 

\begin{figure}
    \centering
    \includegraphics[width=0.49\textwidth]{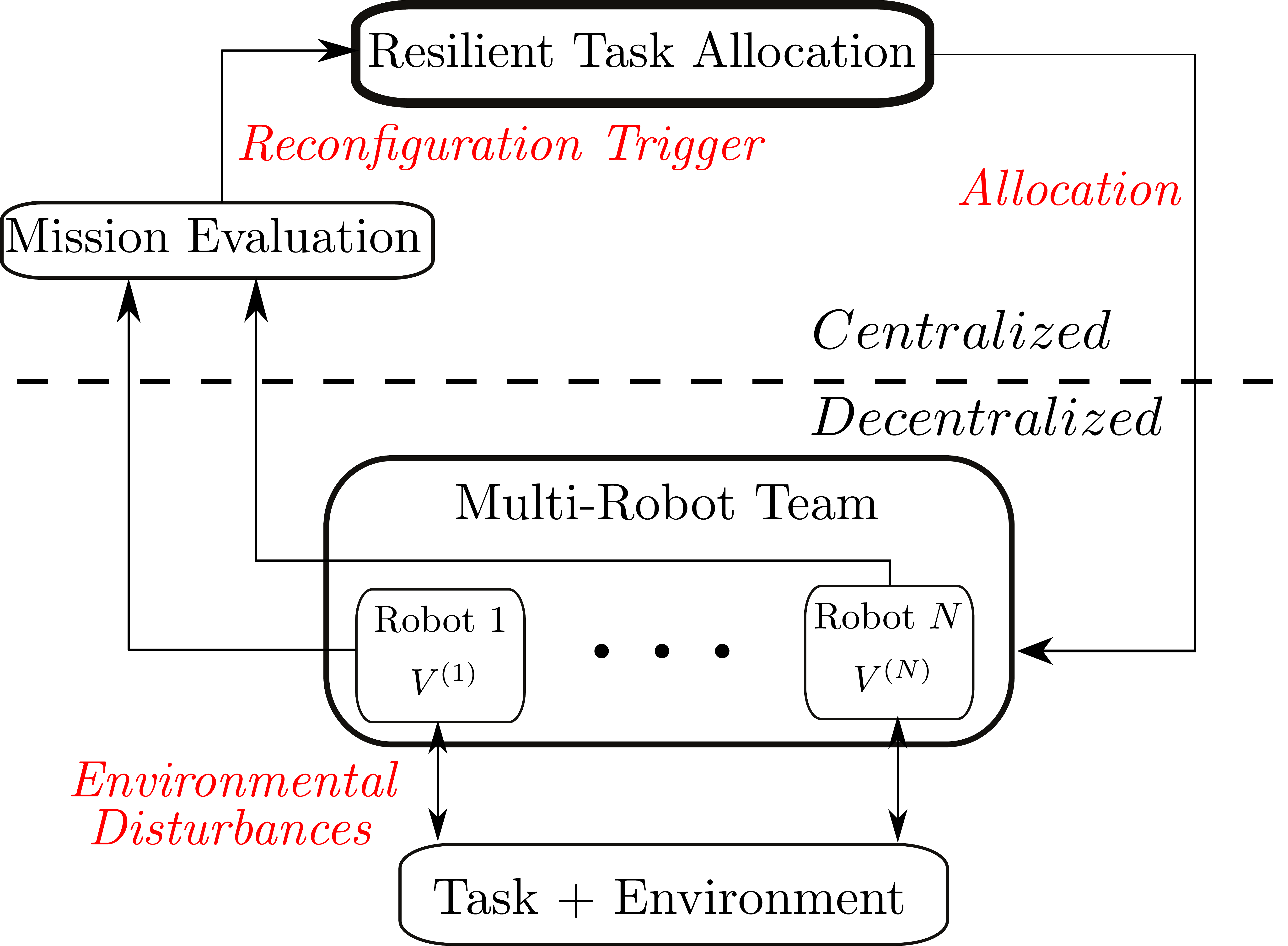}
    \caption{Architecture diagram for the resilient task allocation framework presented in this paper. The task performance discrepancies computed by the robots (using the \ch{task-value functions $V^{(i)}$}) are converted into capability degradation scores in a centralized fashion by the mission evaluation block. If a sufficiently large change in performance is detected, the resilient task allocation algorithm is invoked to redistribute robots among tasks while taking into account their degraded capabilities.}
    \label{fig:miqp_comp}
\end{figure}

To circumvent the computationally intensive nature of solving MIQPs frequently, we present an event-triggered execution framework, where the MIQP is solved only when the estimated capability degradations change beyond a certain threshold. Figure~\ref{fig:miqp_comp} illustrates the system architecture for the resilient task allocation paradigm presented in this paper. $V^{(i)}$ denotes the task-value function associated with robot $i$, which is used to compute the difference between the measured and predicted performance of the robots. This information is used to compute degradation metrics for the different robot capabilities by the mission evaluation block which decides if a reallocation of robots to tasks is warranted or not.  
\section{TASK PERFORMANCE EVALUATION} \label{sec:task_perf}
In this section, we first characterize the heterogeneity within the robot team in terms of the different types of robots available, and the capabilities possessed by each type of robot. This framework is then coupled with a task execution model to quantify the extent to which robot capabilities are affected by environmental disturbances within each task.
\subsection{Robot Heterogeneity} \label{subsec:hetero}
We consider a team of $N$ heterogeneous robots, indexed by the set $\mc{R} = \{1,\ldots,N\}$. Let $U$ denote the total number of unique task-related capabilities available to the robot team, e.g., perception, ground mobility, aerial mobility, object manipulation, etc. Individual robots can exhibit different combinations of capabilities, depending on their size, power, and cost constraints. \ch{These capabilities can be either cumulative---modeled as summable continuous quantities; or non-cumulative---modeled as binary values associated with meeting certain minimum requirements, similar to the approach in~\cite{ravichandar2020strata}.}  \par 
In the literature, robots with identical sets of capabilities are often said to belong to the same \emph{species}~\cite{prorok2016formalizing}. Let $S$ denote the total number of species in the team. Let $\sy{cap}\in \ch{\mathbb{R}_{+}}^{S\times U}$ denote the \emph{capability matrix}, which specifies the capabilities available to each robot species:
\begin{equation}
    \sy{cap} = \begin{bmatrix}
    \mb{q}^{(1)} & \mb{q}^{(2)} & \ldots & \mb{q}^{(S)}
    \end{bmatrix}^T,
\end{equation}
where $\mb{q}^{(s)} = [q^{(s)}_1,\ldots,q^{(s)}_U]^T \in \ch{\mathbb{R}_{+}}^{U}$ is a vector describing the capabilities available to species~$s$. Section~\ref{sec:ctt} and various examples throughout the paper will demonstrate how physically meaningful values can be assigned to the robot capabilities. Let $\sy{cap_bin}\in \{0,1\}^{S\times U}$  denote the binary version of $\sy{cap}$, where, $ \sy{cap_bin}_{su} = 1$ if and only if $q_u^{(s)} > 0$.
Similarly, let $\mb{P} \in \{0,1\}^{S\times N}$ denote the \emph{robot-species mapping matrix}, whose binary-valued element $\mb{P}_{si} = 1$ if and only if robot $i$ belongs to species $s$. \ch{A robot is only allowed to belong to one species, implying that the columns of matrix $\mb{P}$ always sum to 1.}
\subsection{Task Execution}
We now introduce a model for the execution of different tasks by the multi-robot team. The following sections will leverage this model to allow each robot to evaluate its performance at a given task. Let $M$ denote the total number of tasks among which the robots must be allocated. Subsequently, let $\mc{T}_m \subseteq \mc{R}$ represent the index set of robots that are currently allocated to task $m \in \{1,\ldots,M\} \coloneqq \mc{M}$. We assume that robots can only contribute to one task at a time, so $\mc{T}_m \cap \mc{T}_n = \emptyset, \forall m \neq n \in \mc{M}$. We let $x_i \in \mathbb{R}^p$ denote the state of robot $i\in\mc{R}$, and $u_i \in \mathbb{R}^q$ denote the control input, which modifies the state according to the following control-affine dynamics:
\begin{equation} \label{eqn:rob_dyn}
    \dot x_i = f(x_i) + g(x_i)u_i + \ch{D_i(x_i,t)},
\end{equation}
\ch {where $D_i(x_i,t)$ is an unknown term representing the time-varying environmental disturbances acting on the robots. Characterizing $D_i$ can be difficult as it is based on the nature of the disturbance and its interactions with the different capabilities of the robots. Consequently, we directly measure and characterize the performance of the robots in the tasks.} \par 
\ch{In this paper, we encode the current state of each multi-robot task via a non-negative scalar, which we call as the \emph{task-value} function. A large class of robotic tasks can be encoded in this manner---e.g., when robots modify their states according to the gradient flow of such a functional~\cite{turpin2014capt,cortes2017coordinated} or when the task-value function directly represents a quantity relevant to the task~\cite{burgard2000collaborative}. To this end, let $V_m: \mathbb{R}^{p|\mc{T}_m|} \rightarrow \mathbb{R}$ denote the task-value function corresponding to task $m \in \mc{M}$. We assume that this function can be expressed as the composition of robot-wise task-value functions,
\begin{equation} \label{eq:cost}
    V_m(\mb{x}_m) = \bigoplus_{i\in\mc{T}_m} V_m^{(i)}(\mb{x}_m),
\end{equation}
where $\mb{x}_m \in \mathbb{R}^{p|\mc{T}_m|}$ represents the stacked ensemble state of robots allocated to task $m$, and $|\cdot|$ denotes the set cardinality operator. The composition operator $\bigoplus$ could represent operations like summation, products, or minimization, depending on the task.}
Note that, the individual task-value function $V_m^{(i)}$ in~\eqref{eq:cost} can depend on the states of other robots in the task, as is common in coordinated control multi-robot tasks~\cite{cortes2017coordinated}.

\subsection{Task Performance Discrepancy}
As discussed in Section~\ref{sec:intro}, we would like to endow the robots with an ability to evaluate their performance in the tasks, with the aim of quantifying the extent of degradation of different robot capabilities within each task. \ch{We assume that deviations in the evolution of the task-value function are necessarily caused by disturbances introduced in~\eqref{eqn:rob_dyn} and not by external factors.} \par 
Towards this end, we allow each robot in task~$m$, $i\in\mc{T}_m$, to compute a predicted \ch{task-value function} $^{pred}V_m^{(i)}$ which represents its value at the next time step.
More specifically, we consider discrete time intervals, indexed by $t\in\mathbb{N}$, and evenly spaced by a small time interval $\Delta t$, at which the predicted \ch{task-value function} is computed as,
\begin{equation}\label{eq:sim_cost}
^{pred}V_m^{(i)}[t+1] = V_m^{(i)}(\mb{x}_m[t]) + \Delta t \frac{d V_m^{(i)}(\mb{x}_m[t])}{d t}
\end{equation}
where,
\begin{multline} \label{eq:cost_dot}
    \frac{d V_m^{(i)}(\mb{x}_m[t])}{d t} = \frac{\partial V_m^{(i)}(\mb{x}_m[t])}{\partial x_i}\dot{x_i} + \\ \sum_{r\in\mc{N}_i} \frac{\partial V_m^{(i)}(\mb{x}_m[t])}{\partial x_r}\dot{x_r}.
    \end{multline}

Here, $\mc{N}_i$ represents the neighborhood set of robot $i$, and can be described using a graph embedding---for example, representing physical proximity among the robots~\cite{cortes2017coordinated}. Some examples of multi-robot tasks described in this fashion include coverage control~\cite{cortes2004coverage}, formation control~\cite{oh2015survey}, rendezvous~\cite{yan2013survey}, and target tracking~\cite{pimenta2009simultaneous}. \par 

At discrete time $t$, robot $i$ can then use~\eqref{eq:sim_cost}, to compute the predicted \ch{task-value} at time $t+1$. Comparing this against the measured \ch{task-value} function at the next time step allows the robot to evaluate its task performance as discussed next. 
\begin{definition}[Task Performance Discrepancy] \label{defn:task_disc}
Let $\Delta V^{(i)}[t+1]$ denote the discrepancy associated with the task performance of robot $i$ at time $t+1$, given as,
\begin{multline} \label{eq:task_disc}
    \Delta V^{(i)}[t+1] = \\
     \min\left\{\max\left\{1 - \frac{V_m^{(i)}(\mb{x}_m[t+1]) - V_m^{(i)}(\mb{x}_m[t])}{^{pred}V_m^{(i)}[t+1] - V_m^{(i)}(\mb{x}_m[t])},0\right\},1\right\}.
\end{multline}
For a small time interval $\Delta t$, the task-performance discrepancy $\Delta V^{(i)}$ encodes the fractional deviation between how much progress the robot made towards modifying its \ch{task-value function} (encoded in the numerator) and how much progress it expected to make in the same time interval (encoded in the denominator). 
\end{definition}
As seen in Definition~\ref{defn:task_disc}, if the robot did not experience any disturbance, the predicted \ch{task-value} $^{pred}V_m^{(i)}[t+1]$ and the actual measured task-value $V_m^{(i)}(\mb{x}_m[t+1])$ would be equal \ch{(barring approximation errors and variations caused due to non-idealities in the motion of the robots)}. This would imply that the task performance discrepancy $\Delta V^{(i)}[t+1]$ would be equal to (or close to) 0. Similarly, a discrepancy value of~$1$ implies that the robot did not make any progress towards the task execution. As seen in~\eqref{eq:task_disc}, we cap values of $\Delta V^{(i)}$ which are less than~$0$ or greater than~$1$, which correspond to situations where the robot did better than expected, or its actions resulted in an unexpected direction of change of the \ch{task-value} function, respectively.
In the following example, we demonstrate how the task performance discrepancy can quantify the real-time disturbances experienced by a multi-robot system.
\begin{example} \label{ex:task_disc}
Consider a multi-robot team composed of two robots: a ground ``leader" robot $r_1$ and an aerial ``follower" robot $r_2$. The ground robot is tasked with tracking a moving goal and the aerial robot is tasked with maintaining a pre-specified distance with respect to the ground robot. The robot-wise \ch{task-value functions}, whose minimization encodes these objectives, are given as,
\begin{align}
    & V^{(1)} = 0.5\|x_1 - g\|^2 \\
    & V^{(2)} = 0.5(\|x_2 - x_1\| - d)^2,
\end{align}
where $g$ represents the location of the goal and $d$ represents the desired following distance for the aerial robot. For simplicity, we model the motion of both robots using single integrator dynamics: $\dot x = u$. At time $t=0.66s$, gusts of head wind affect the motion of the aerial robot, but not the ground robot. We model this disturbance as a multiplier to the control input applied by the robot. More specifically, for robot $r_2$, $\dot x_2 = (1-w)u_2$ where $w$ gradually increases from $0$ to $0.3$. Figure~\ref{fig:ex1:b} shows how the task performance discrepancy corresponding to both the robots, evolves. The higher values of discrepancy for robot $r_2$ capture the fact that, the robot is making a smaller amount of progress towards minimizing its cost function than it expects, as computed by~\eqref{eq:task_disc}. 
\begin{figure}
    \centering
    \subfloat[][]{
    \includegraphics[width=0.245\textwidth,clip, trim={1cm 0.0cm 1cm 1cm}]{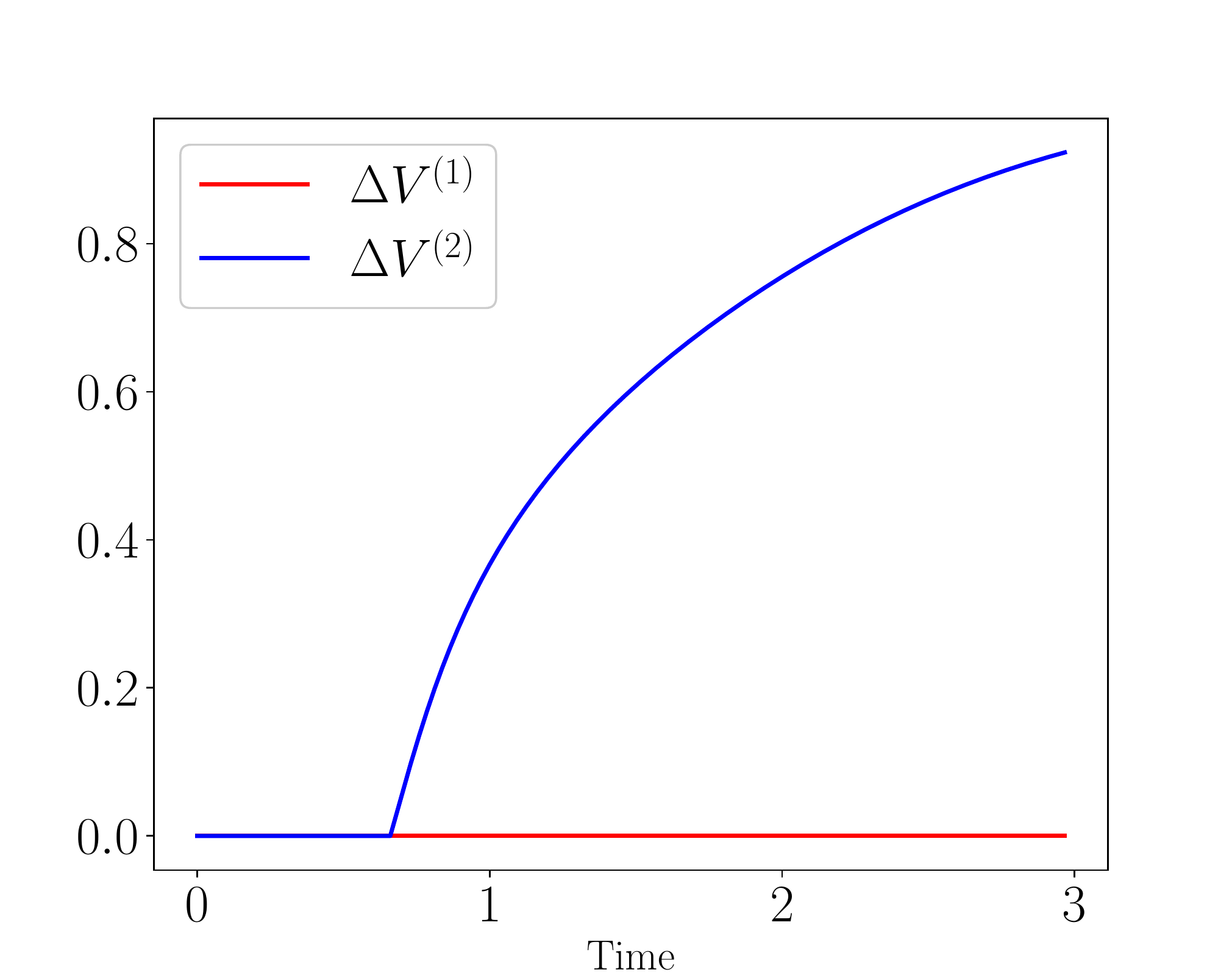}
    \label{fig:ex1:b}}
    \subfloat[][]{
    \includegraphics[width=0.240\textwidth,clip, trim={1cm 0.0cm 1cm 1cm}]{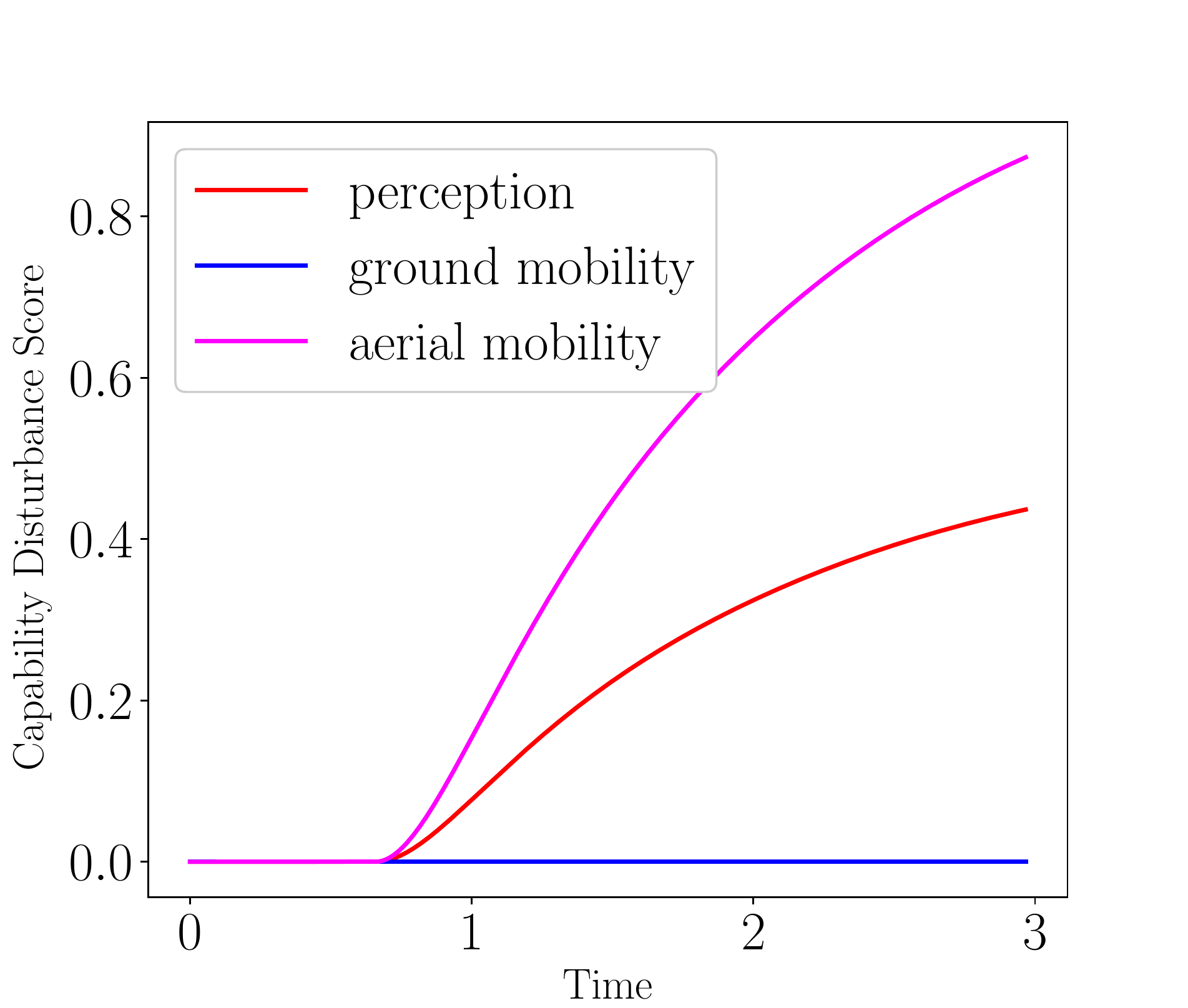}
    \label{fig:ex2:a}}
    \caption{Task-performance discrepancy and capability degradation metrics for a goal tracking task executed by a ground and aerial robot, $r_1$ and $r_2$, respectively. As described in Example~\ref{ex:task_disc}, at time $t=0.66s$, the aerial robot experiences a simulated wind disturbance. Figure~\ref{fig:ex1:b} illustrates a corresponding increase in the task performance discrepancy computed by the robot, according to~\eqref{eq:task_disc}. As explained in Section~\ref{subsec:cap_dis} and in Example~\ref{ex:cap_dist}, this causes an increase in the capability degradation of aerial mobility (see Fig.~\ref{fig:ex2:a}).}
    \label{fig:example_task_disc}
\end{figure}
\end{example}
\subsection{Capability Degradation Metrics} \label{subsec:cap_dis}

While \eqref{eq:task_disc} gives us the robot-wise task performance discrepancies, it does not tell us which \emph{robot capabilities} are affected by environmental disturbances. Towards that end, we assemble the task performance discrepancies of the robots in task $m\in\mc{M}$ into a vector denoted as $\mb{\Delta V}_m \in [0,1]^{|\mc{T}_m|}$. In the following definition, we use the heterogeneous mappings described in Section~\ref{subsec:hetero} to compute a capability degradation metric for the robots in each task.
\begin{definition}\label{defn:cap_dist}
Let $\mb{d}_m^*[t] \in [0,1]^U$ denote the extent to which each capability is degraded within task $m$ at time $t$. The higher the score, the more ineffectual the robots having this capability are at executing task $m$. We compute this capability degradation metric based on the task performance discrepancy values computed in~\eqref{eq:task_disc},
\begin{equation}\label{eq:cap_dist_inst}
    \mb{d}_m^*[t] = \mb{\overline Q}_{S_m,-}^T\mb{P}_{S_m,\mc{T}_m}\mb{\Delta V}_m[t] ,
\end{equation}
where $\mb{P}_{S_m,\mc{T}_m}$ denotes a submatrix of $\mb{P}$ which contains only the rows and columns corresponding to the species and indices of robots currently present in task $m$, respectively. $\mb{\overline Q}_{S_m,-}$ contains the rows corresponding to the species of robots in task $m$ along with all columns. The rows of $\mb{P}_{S_m,\mc{T}_m}$ and columns of $\mb{\overline Q}_{S_m,-}$ are normalized to preserve the value of the disturbances between 0 and 1. 
\end{definition}

Note that~\eqref{eq:cap_dist_inst} represents the instantaneous capability degradation at time $t$ based on the task performance discrepancies $\mb{\Delta V}_m[t]$. We introduce the following update law to capture a time-averaged version of the capability degradation metrics,
\begin{equation} \label{eq:dist_score_update}
    \mb{d}_m[t+1] = \mb{d}_m[t] + \Delta t\mb{\Theta}_m[t]\Big(\mb{d}_m^*[t] - \mb{d}_m[t]\Big) ,
\end{equation}
where $\mb{d_m}[t]$ now represents the time-averaged capability degradation at discrete time $t$. Here, $\mb{\Theta}_m$ is a binary diagonal matrix, whose $u^{\text{th}}$ diagonal element indicates whether capability $u$ is currently available on any robot allocated to task $m$, defined as,
\begin{equation} \label{eq:cap_indicator}
    \mb{\Theta}_m[t] = \mathrm{diag}\left(\mb{1}^T\mb{\overline Q}_{S_m,-} \right)
\end{equation}
where for $g\in\mathbb{R}^n$, $\mathrm{diag}(g) = G \in \mathbb{R}^{n\times n}$ and the columns of $\mb{\overline Q}_{S_m,-}$ are normalized as before. The introduction of $\mb{\Theta}_m$ allows us to update only the degradation values for the capabilities which are currently deployed in task $m$, and keep the other values constant.
The following example continues the scenario presented in Example~\ref{ex:task_disc} to illustrate how the update law presented in~\eqref{eq:dist_score_update} can be leveraged.
\begin{example} \label{ex:cap_dist}
For the heterogeneous multi-robot team presented in Example~\ref{ex:task_disc}, we first specify the capability matrix $\mb{Q}$, which consists of three capabilities---perception (measured in terms of the area that the robot can sense around it), ground mobility, and aerial mobility (both measured in terms of speed):
\begin{equation}
    \mb{Q} = \begin{bmatrix}
    10~\emph{m}^2 & 2~\emph{m/sec} & 0~\emph{m/sec} \\ 
    10~\emph{m}^2 & 0~\emph{m/sec} & 5~\emph{m/sec} \\ 
    \end{bmatrix}.
\end{equation}
The robot-species mapping is simply: $\mb{P} = \mathbb{I}_2$, and $\mb{\Theta}=\mathbb{I}_3$ since all three capabilities are present in the task. For the same scenario presented in Example~\ref{ex:task_disc}, Fig.~\ref{fig:ex2:a} plots each element of the capability degradation metric $\mb{d}$ computed according to the update law presented in~\eqref{eq:dist_score_update} (note that the task index is hidden). As seen, the degradation metric for aerial mobility increases as the task performance discrepancy for the aerial robot is mapped to the capabilities it possesses using~\eqref{eq:cap_dist_inst}. The degradation metric for the perception capability also increases, where the lower magnitude is explained by the fact that it represents the average degradation experienced in this capability by the ground and the aerial robot (the former of which is unaffected by the wind).  
\end{example}
\section{RESILIENT TASK ALLOCATION} \label{sec:task_all}
In this section, we develop an optimization-based task allocation framework which meets the resilience objectives described in Section~\ref{sec:intro}. Towards this end, we take into account the fact that, tasks can often be accomplished with one of \emph{multiple possible capability configurations}. For example, a surveillance task over a large region could be accomplished by slow moving ground robots with large perception ranges, or fast moving aerial robots with smaller perception ranges. This notion is formalized in the definition below. 
\begin{definition}[Task Requirement Matrix] \label{defn:task_req}
Let $K_m$ denote the number of possible alternative configurations of capabilities which can support the accomplishment of a given task $m \in \mc{M}$. We denote $\sy{task_req}:\mathbb{R}_{\geq 0}^{K_m \times U}$ as the requirement matrix for task $m$, which specifies the aggregated capabilities required to effectively execute the task in each of the different configurations. In other words, each row of $\sy{task_req}$ specifies a possible combination of minimum aggregated capabilities which need to be assigned to task $m$.
\end{definition}
In this paper, we are interested in generating an \emph{allocation matrix},  $\mb{A} \in \{0,1\}^{M \times N}$, whose element $\mb{A}_{ji} = 1$ if and only if robot $i$ is allocated to task $j$. For each task $m\in\mc{M}$ and candidate allocation $\mb{A}$, let $\mb{c}_m \in \mathbb{R}_{+}^{U}$ denote the total aggregated capabilities assigned to the task (computed in a similar manner to~\cite{ravichandar2020strata}), given as,
\begin{equation} \label{eq:ideal_capab}
    \mb{c}_m = \left(\mb{A}_{m,-}\mb{P}^T\mb{Q}\right)^T,
\end{equation}
where $\mb{A}_{m,-}$ denotes the $m^{th}$ row of $\mb{A}$. However, as discussed in Section~\ref{sec:task_perf}, the performance of different robot species will be different in the tasks, due to environmental disturbances. To explicitly account for these variations in the allocation process, we introduce the \emph{effective} total capabilities assigned to a given task, which leverages the capability degradation metrics computed in~\eqref{eq:dist_score_update}. Thus, the effective aggregated capabilities in task $m$ can be given as,
\begin{equation} \label{eqn:eff_capabilities}
    \mb{\hat c}_m =  \mb{c}_m - \left(\mb{d}_m \odot  \mb{c}_m\right),
\end{equation}
where $\odot$ is the Hadamard product. Using~\eqref{eqn:eff_capabilities}, the following definition outlines the conditions which would ensure that a sufficient amount of aggregated capabilities are assigned to the tasks. 
\begin{definition}[Effective Task Execution] \label{defn:capab_margin}
The capability requirements for a given task $m \in \mc{M}$ are met, when the effective aggregated capabilities of the robots allocated to it are greater than those specified by one or more of the configurations in the task requirements matrix (see Definition~\ref{defn:task_req}). This is encoded by the following two conditions:
\begin{align}
&\mb{\hat c}_m - \left(\boldsymbol{\iota}_m^T\mb{Y}_m^*\right)^T = \boldsymbol{\delta}_m \\ 
& \boldsymbol{\delta}_m \geq 0 \label{eq:delta_perfect}
\end{align}
where $\geq 0$ is interpreted element-wise, and $\boldsymbol{\iota}_m \in \{0,1\}^{K_m}$ is an indicator matrix specifying which of the $K_m$  possible configurations is selected. In particular, the condition $\boldsymbol{\iota}_m^T\mb{1} = 1$ ensures that only one configuration is selected at a given point in time. $\boldsymbol{\delta}_m \in \mathbb{R}^U$ then represents the aggregated capability margin, which is the difference between the total available capabilities assigned to the task and the requirements of the task.
\end{definition}
However, environmental conditions might force a situation where it is impossible to meet the requirements for every task, i.e. constraint \eqref{eq:delta_perfect} might not be satisfied $\forall m\in\mc{M}$. To impart a second layer of resilience to our framework, we introduce a \emph{task relaxation} matrix $\boldsymbol{\phi}\in \{0,1\}^M$ which indicates whether the requirements for each task will be met or not, 
\begin{equation}
    \boldsymbol{\phi}_m = \begin{cases}
    0, ~ \text{task } m \text{ requirements are being met} \\
    1, ~ \text{otherwise}.
    \end{cases}
\end{equation}
Using $\boldsymbol{\phi}$, we can modify the requirement in~\eqref{eq:delta_perfect} as follows: $\boldsymbol{\delta}_m \geq -\boldsymbol{\phi}_m\delta_{max}\mb{1}$, where $\delta_{max} \in \mathbb{R}$ represents the maximum extent to which the task requirements constraints can be violated for all capabilities.\par 

We now present the mixed-integer quadratic program (MIQP) which can be solved to generate a resilient task allocation for the multi-robot team. \par 
\begin{subequations}\label{eq:allocationalgorithm}
\begin{flalign}
\text{\bf Resilient and Resource-Aware Task allocation } \tag{\ref{eq:allocationalgorithm}}&&
\label{eq:allocationalgorithmactual}
\end{flalign}
\begin{align}
\begin{split}
\minimize_{\substack{\boldsymbol{\iota}_1,\boldsymbol{\iota}_2,\ldots,\boldsymbol{\iota}_M, \\ \mb{A}, \boldsymbol{\phi}}} & \mb{1}^T(\mb{A}\mb{P}^T)\mb{W}_s\mb{1} + \mb{w}_J^T\boldsymbol{\phi} +   \\
&\quad{}\qquad{} l\|\boldsymbol{D}\mb{1}\|_2^2  + \|\mb{1}^T\mb{T}(\mb{A}-\mb{A}_p)\|_1 
\end{split}\label{eq:miqp:a}\\
\subjto & \mb{\hat c}_m - \left(\boldsymbol{\iota}_{m}^T\mb{Y}_m^*\right)^T = \boldsymbol{\delta}_m, \label{eq:miqp:b}\\ 
&\boldsymbol{D} \geq -\boldsymbol{\phi}\mb{1}^T\delta_{max} \label{eq:miqp:c}\\
&\mb{1}^T(\mb{A}\mb{P}^T) \leq \boldsymbol{\lambda}^T \label{eq:miqp:d}\\ 
& \mb{1}^T\mb{A} \leq \mb{1} + \left(\Delta V_{\text{thresh}}\mb{1} - \mb{\Delta V}\right) \label{eq:miqp:e}\\
& \boldsymbol{\iota}_m^T\mb{1} = \mb{1} \label{eq:miqp:f}\\
&\hspace{10em}\forall m \in \mc{M}, \notag
\end{align}
\noeqref{eq:miqp:a}\noeqref{eq:miqp:b}\noeqref{eq:miqp:c}\noeqref{eq:miqp:d}\noeqref{eq:miqp:e}\noeqref{eq:miqp:f}
\end{subequations}
where $\boldsymbol{D} = [\boldsymbol{\delta}_1,\boldsymbol{\delta}_2, \ldots, \boldsymbol{\delta}_M]^T \in \mathbb{R}^{M\times U}$ and the inequality in constraint~\eqref{eq:miqp:c} holds elementwise. We will now define the symbols and the roles played by various terms in the above defined optimization problem. First, $\mb{W}_s \in\mathbb{R}_+^{S \times S}$ is a diagonal weight matrix which represents the cost-of-deployment associated with robots of different species (for instance, a robot with an expensive LIDAR might have a higher cost of deployment associated with it). Furthermore, $\mb{w}_J\in\mathbb{R}_+^{M}$ represents the relative importance among the various tasks, which is taken into account when considering which task constraint to relax first. For example, when considering the mission objective of defending a perimeter~\cite{shishika2019team}, it might be better to relax the constraints of the patrol task (which detects new intruders) than the defense task (which intercepts them) if both cannot be achieved simultaneously. \par
The third term in the cost function~\eqref{eq:miqp:a}, $\|\boldsymbol{D}\mb{1}\|_2^2$, scaled by a positive constant $l$, serves two purposes: it penalizes excessive allocation of capabilities to a given task and also ensures that in case the constraints corresponding to a given task are relaxed due to significant environmental disturbances, they are done so to \ch{the least extent possible (in a Pareto-optimal sense)}.
The final term in the cost function~\eqref{eq:miqp:a}, $\|\mb{1}^T\mb{T}(\mb{A}-\mb{A}_p)\|_1$, represents the cost of transitioning robots between the tasks. In this regard, $\mb{A}_p$ simply represents the current allocation of the multi-robot team to tasks (computed as the solution of the MIQP in the previous iteration, see Algorithm~\ref{alg:task_all}). The transition cost matrix, $\mb{T} \in \mathbb{R}_{+}^{M\times M}$ is a diagonal matrix, where $|\mb{T}_{i,i} - \mb{T}_{j,j}|$ represents the cost incurred by each robot when it transitions from task $i$ to task $j$, or vice versa. Similarly, $\mb{T}_{i,i}$ simply indicates the cost associated with an idle (unallocated) robot being assigned to task $i\in\mc{M}$. For instance, these costs can be assigned by the mission designer based on the distances that robots have to traverse when transitioning between tasks. \par 
The vector $\boldsymbol{\lambda}\in\mathbb{N}^S$ represents the total number of robots of each species available for allocation and thus, constraint~\eqref{eq:miqp:d} ensures that the resource constraints of the overall team are accounted for by the allocation algorithm. Along a similar vein, constraint~\eqref{eq:miqp:e} ensures that each robot is allocated to only one task at most. Here, $\mb{\Delta V} \in [0,1]^N$ represents the stacked task-discrepancies corresponding to the entire team (see Definition~\ref{defn:task_disc}), and $\Delta V_{\text{thresh}}\in [0,1]$ represents the maximum acceptable task performance discrepancy that a given robot can have, for it to be eligible for allocation by the algorithm. Thus, if the following condition holds for robot $i\in\mc{R}$, 
\begin{equation}
 \Delta V_{\text{thresh}} - \Delta V^{(i)} < 0,
\end{equation}
then robot $i$ will not be allocated to any task, since it is deemed unfit to perform any task. For instance, a ground robot stuck in a crevice might not be able to perform any task in the environment, and will not be considered in the allocation. \par 
\subsection{Computational Aspects}
As discussed earlier, the capability degradation metric $\mb{d}_m$ is updated every $\Delta t$ seconds, which is then incorporated into the resilient task allocation optimization problem~\eqref{eq:allocationalgorithm} via constraint~\eqref{eq:miqp:b}. However, if we assume that environmental disturbances affect the multi-robot team at time scales much larger than $\Delta t$, it is clear that the MIQP described by~\eqref{eq:allocationalgorithm} need not be solved every $\Delta t$ seconds. This idea is further reinforced by the fact that, the MIQP must be solved in a centralized manner, and is not amenable to real-time solutions due to its NP-complete nature~\cite{del2017mixed}. \par 
Indeed, a reallocation of robots to tasks is warranted only when there are significant changes in the capability degradation metrics associated with any of the tasks. Let $t_l$ denote the time index when the MIQP was most recently solved. We introduce a binary variable $\beta[t]$, which determines if the MIQP should be solved at time $t$,
\begin{equation} \label{eq:miqp_trigger}
    \beta[t] = \begin{cases}
    1, \text{ if } \exists ~m\in\mc{M}, \text{ s.t. } \max{\left(\mb{d}_m[t] -\mb{d}_m[t_l] \right)} \geq \ch{\chi}, \\
    0, \text{otherwise}
    \end{cases}
\end{equation}
where \ch{$\chi$} is a user defined threshold on the change in any capability degradation value.
\begin{algorithm}
	\caption{Resilient task allocation via online task performance evaluation}
	\label{alg:task_all}
	\begin{algorithmic}[1]
		\Require
		\Statex Robot team heterogeneity specifications $\mb{Q}, \mb{S}, \mb{P}$, $\boldsymbol{\lambda}$
		\Statex Task Specifications $\mb{Y}_m^*$, $\mb{T}$
		\Statex Parameters $\mb{W}_s$, $\mb{w}_J$, $\Delta V_{\text{thresh}}$, $l$, $\boldsymbol{\delta}_{max}, \chi$
		\State Initialize: $t = t_l = 0$, $\mb{A}_p = \mb{0}^{M\times M}$
		\State Compute $\mb{A}$ and transmit to robots. \Comment \eqref{eq:allocationalgorithm}
		\State Set $\mb{A}_p = \mb{A}$
		\While{true}
		\State Execute tasks $m\in\mc{M}$
		\State Compute $\Delta V^{(i)}$, $\forall i\in\mc{R}$\label{step:task_disc}\Comment \eqref{eq:task_disc}
		\State Update capability degradation $\mb{d}_m$, $\forall m\in\mc{M}$\label{step:cap_dist} \Comment \eqref{eq:dist_score_update}
		\State Compute reallocation trigger $\beta$ \Comment \eqref{eq:miqp_trigger}
		\If {$\beta = 1$}
		\State Compute $\mb{A}$ and transmit to robots \Comment \eqref{eq:allocationalgorithm}
		\State Set $\mb{A}_p = \mb{A}$ and $t_l = t$
		\EndIf
		\State $t = t + 1$
		\EndWhile
	\end{algorithmic}
\end{algorithm}
Algorithm~\ref{alg:task_all} outlines the operations of the resilient task allocation framework. Step~\ref{step:task_disc}, \ch{signified by the ``Multi-Robot Team" block in Fig.~\ref{fig:miqp_comp}}, computes the task performance discrepancy values $\Delta V^{(i)}$ based on the environmental disturbances experienced by the robots.  Following this, step~\ref{step:cap_dist} computes the capability degradation metrics $\mb{d}_m$ for each task $m\in\mc{M}$ and uses this to compute $\beta$ using~\eqref{eq:miqp_trigger}. \ch{These operations are represented by the ``Mission Evaluation" block in Fig.~\ref{fig:miqp_comp}}. If $\beta = 1$, the task allocation MIQP presented in~\eqref{eq:allocationalgorithm} is solved to generate the allocation matrix $\mb{A}$ which subsequently results in \ch{a} rearrangement of robots among the tasks. \ch{These steps require the robots to communicate their task discrepancies to the central coordinator, and receive the robot-task allocations.}
The next section illustrates the salient features of the proposed framework in a heterogeneous multi-robot coverage control and target tracking scenario.
\section{ENVIRONMENT COVERAGE AND TARGET TRACKING: AN APPLICATION} \label{sec:ctt}
\begin{figure}
\centering
\subfloat[][]{\label{subfig:exp:4}\includegraphics[width=0.23\textwidth]{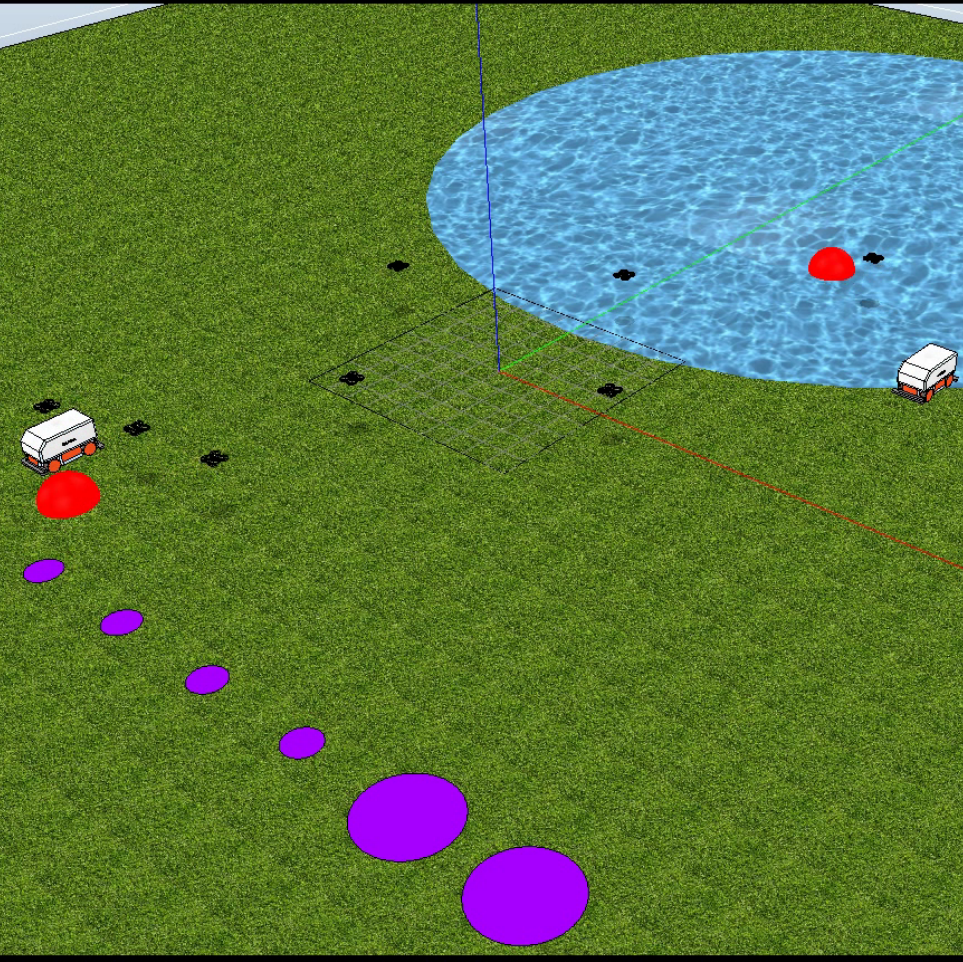}}~
\subfloat[][]{\label{subfig:exp:1}\includegraphics[width=0.23\textwidth]{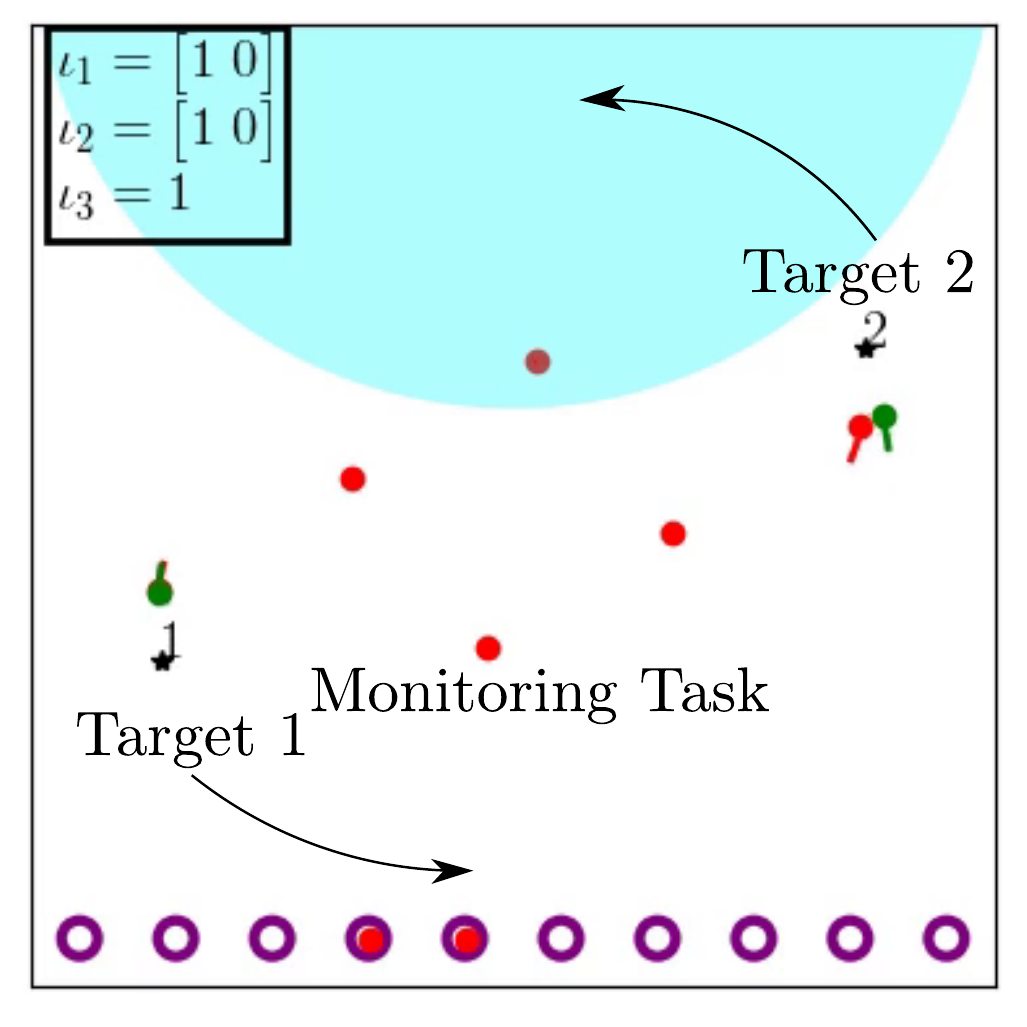}}\\
\subfloat[][]{\label{subfig:exp:2}\includegraphics[width=0.23\textwidth]{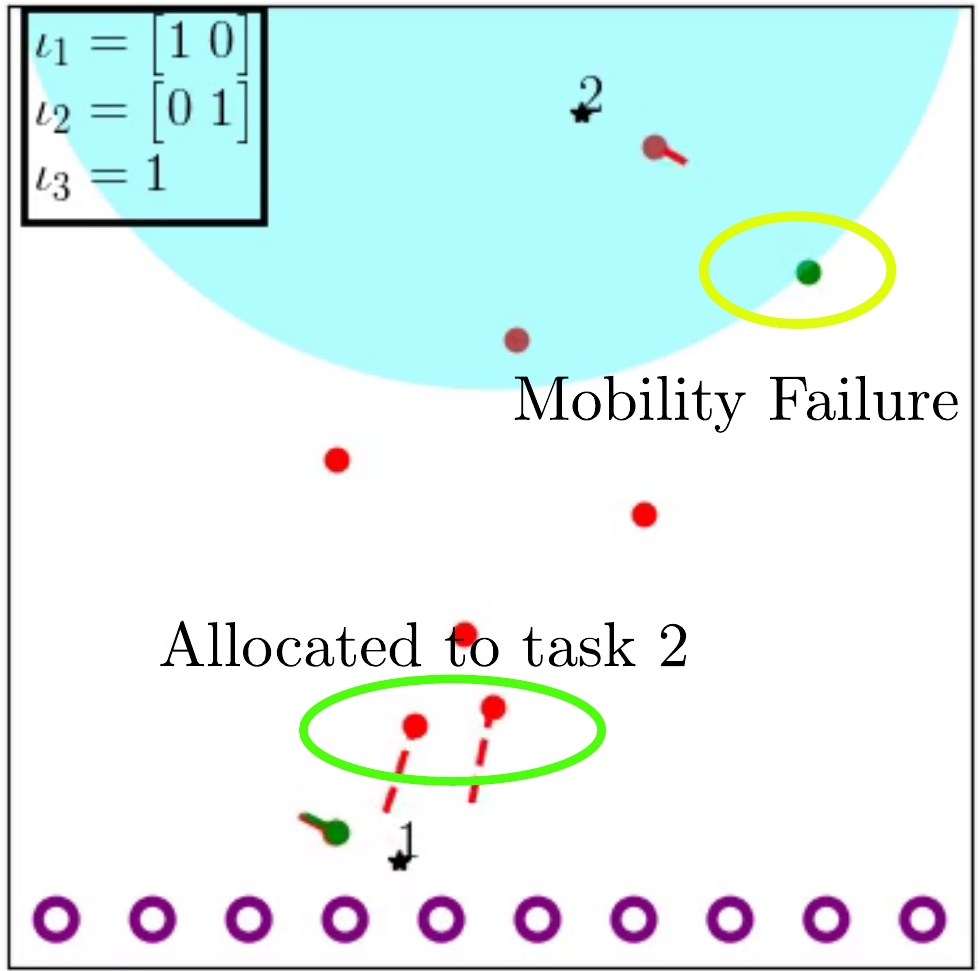}}~
\subfloat[][]{\label{subfig:exp:3}\includegraphics[width=0.23\textwidth]{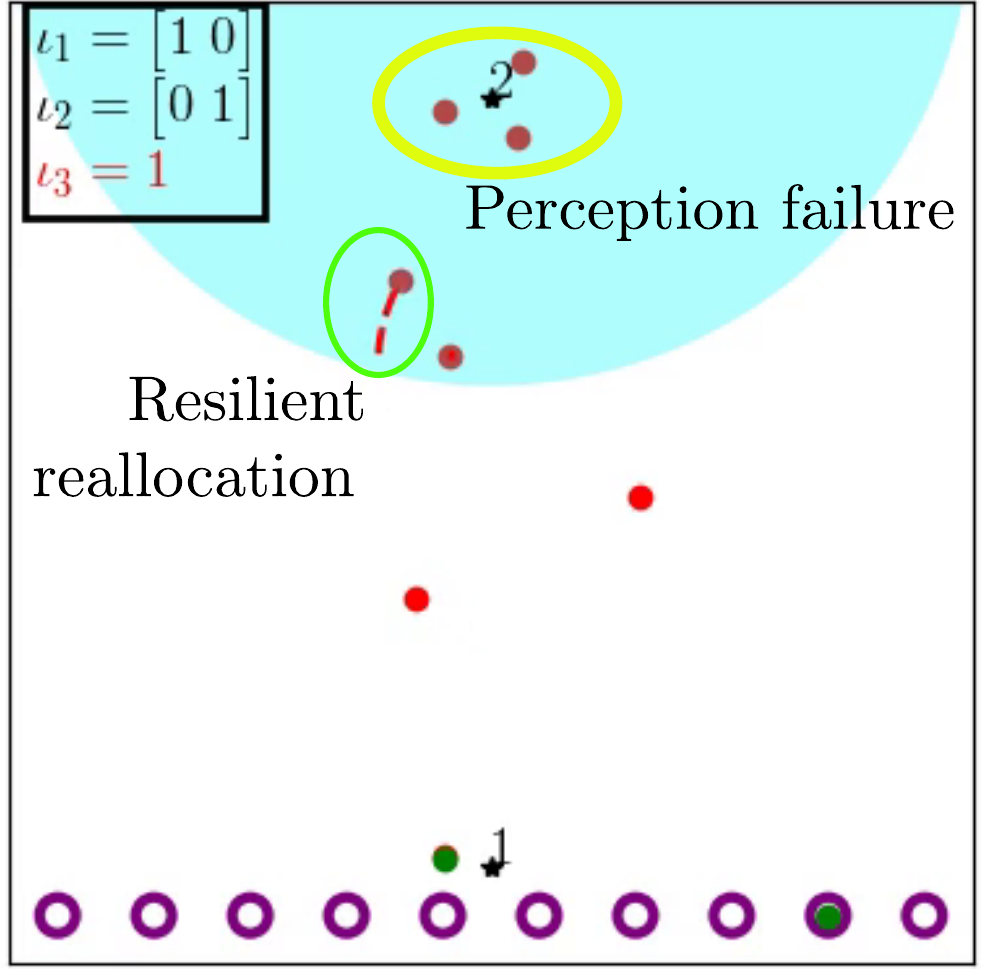}}~

\caption{Resilient task allocation experiment. \ref{subfig:exp:4}: Simulation scenario in CoppeliaSim. In~\ref{subfig:exp:1}-\ref{subfig:exp:3}, aerial and ground robots are denoted by red and green circles respectively; they are allocated among three tasks: tracking two different targets, and monitoring the environment.  In~\ref{subfig:exp:2}, the ground robot tracking target 2 becomes immobilized after encountering a (blue) low friction surface (shown by the yellow ellipse). The proposed algorithm identifies the ground mobility capability as being compromised and deploys two aerial robots to join the task instead (highlighted by the green ellipse), thus mitigating the effects of the failure. In~\ref{subfig:exp:3}, a perception failure is depicted near task 2 requiring the presence of additional robots. The allocation algorithm autonomously chooses to sacrifice performance in the monitoring task (encoded as a lower priority task) and redirects a robot accordingly (highlighted by the green ellipse in~\ref{subfig:exp:3}). }
\label{fig:exp_scenario}
\end{figure}

\ch{As illustrated in Fig.~\ref{fig:exp_scenario}, we consider a team of aerial and ground robots (simulated in CoppeliaSim~\cite{coppeliaSim}), which need to be allocated among three tasks: tracking target 1 (task 1), tracking target 2 (task 2), and monitoring of the environment (task 3).} In particular, we use the coverage control algorithm~\cite{cortes2004coverage} to execute the monitoring task, with the importance density function chosen as a zero-centered Gaussian function. The robots performing tracking tasks $1$ and $2$ are also required to maintain a certain quality of surveillance on the target, which is modeled as a function of both the distance to the target and a scalar state $e_i$ denoting the environmental effects on the sensing. These task objectives are encoded into the following cost functions whose minimization represents the execution of the tasks, 
\begin{align}
    &V_k = \frac{1}{2}\sum_{i\in\mc{T}_k} (\|x_i - \gamma_k\|^2 - d_{k})^2+e_i\|x_i- \gamma_k\|^2 + \\ 
    &\qquad \qquad\left(\sum_{j\in\mc{T}_k \setminus {i}}\frac{1}{\|x_i - x_j\|_2^2} - \frac{1}{d_0^2}\right)^2, k = 1,2,\nonumber \\
    &V_3 = \frac{1}{2}\sum_{i\in\mc{T}_3} \|x_i - c_i\|^2,
\end{align}
where $\gamma_k$ denotes the locations of target $k$, $d_k$ denotes the desired distance to be maintained between the robots and target $k$, $d_0$ determines the minimum distance maintained between the robots in the task, and $c_i$ denotes the \ch{desired location to drive to for robot $i$, as determined by the coverage control algorithm}~\cite{cortes2004coverage}. \par 
The simulated experiment considers two species of robots: an aerial and a \ch{omnidirectional} ground platform.  The heterogeneity among the robots is characterized via five capabilities: perception (m$^2$), sensing resolution (m), airspeed (m/sec), ground speed (m/sec), and communication rate (Mb/sec). These specifications are captured by the robot capability matrix:
\begin{equation}
    \mb{Q} = \begin{bmatrix} 
    5~\text{m}^2 & 1~\text{m} & 3~\text{m/sec} & 0~\text{m/sec} & 5~\text{Mb/sec} \\ 
    2~\text{m}^2 & 3~\text{m} & 0~\text{m/sec} & 1~\text{m/sec} & 8~\text{Mb/sec}
    \end{bmatrix}.
\end{equation}
The task requirements matrix (see Definition~\ref{defn:task_req}) is given as,
\begin{align}
    &\mb{Y}_{1}^* = \mb{Y}_{2}^* = \begin{bmatrix}
    7 & 4 & 3 & 1 & 10 \\
    10.5 & 2.1 & 6.3& 0& 10.5
    \end{bmatrix} \label{eqn_task_1_2_req} \\
    &\mb{Y}_{3}^* = \begin{bmatrix}
    20 & 4 & 12 & 0 & 15
    \end{bmatrix}\label{eqn_task_3_req}.
\end{align}
As seen, the tracking tasks can either be accomplished using a team of aerial and ground robots (configuration 1) or only aerial robots (configuration 2). 
The parameters for the optimization program are chosen as follows: $\mb{w}_J = [100,100,10]$, (indicating that the coverage task is less critical to mission success compared to the target tracking tasks), $\mb{W}_s = \ch{\mathrm{diag}}([0.1,0.1])$, where $\mathrm{diag}$ is the diagonalization operator, \ch{$\Delta V_{\text{thresh}} = 0.9$}, $l = 1.0$, $\delta_{max} = 1000$, \ch{$\chi = 0.33$}, and $\mb{T} = \ch{\mathrm{diag}}([65,18,45])$. \par 
 \ch{The velocities generated by the task controllers were translated into rotor and wheel commands (for the aerial and ground robots respectively), and simulated using the Bullet physics engine~\cite{coumans2013bullet}. It should be noted that the task discrepancies of the robots, even in nominal conditions, were non-zero due to the realistic dynamics of the robots and the first-order approximation used to compute the predicted task-value function in~\eqref{eq:sim_cost}. However, the time-averaged computation of the capability degradation in~\eqref{eq:dist_score_update}, along with the flexibility of adjusting the trigger thresholds $\chi$ and $\Delta V_{\text{thresh}}$, enabled the resilient task allocation algorithm to reject the process noise}. \par
\begin{figure}[h]
    \centering
    \includegraphics[width=0.49\textwidth]{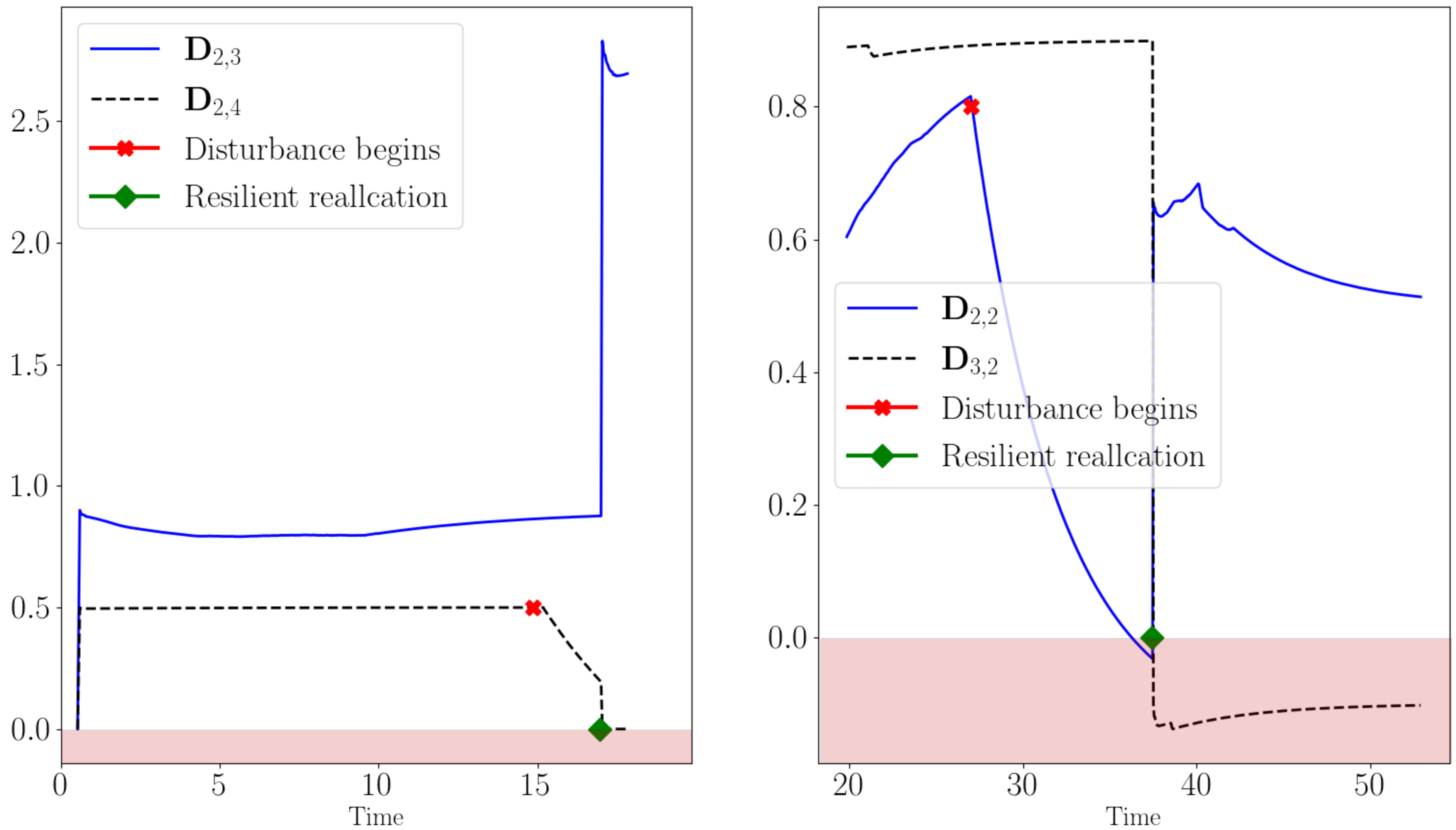}
    \caption{(Left) Margins for aerial and ground mobility capabilities corresponding to tasks 2 ($\mb{D}_{2,3}$ and $\mb{D}_{2,4}$, respectively). As seen, the mobility failure of the ground robot in Fig.~\ref{subfig:exp:2} is captured by a decrease in $\mb{D}_{2,4}$. A value below 0 (indicated by the shaded area) indicates that the requirements for the task are not met anymore. This prompts the allocation algorithm to switch to a different configuration, thus mitigating the failure. (Right) Margins for the sensing resolution capability corresponding to tasks 2 and 3 ($\mb{D}_{2,2}$ and $\mb{D}_{3,2}$, respectively). Due to the environmental disturbance in task 2 (see Fig.~\ref{subfig:exp:3}), the capability margin $\mb{D}_{2,2}$ drops. The resilient allocation algorithm reallocates a robot from task 3 to task 2, ensuring that $\mb{D}_{2,2}$ remains above 0, while intentionally sacrificing performance in task 3.}
    \label{fig:cap_margins}
\end{figure}

 
 \ch{In Figure.~\ref{fig:exp_scenario}, we represent a simplified schematic representation of the experiment.} The red circles represent aerial robots, and the green circles represent the ground robots. Figure.~\ref{subfig:exp:1} shows the initial deployment of robots to tasks as generated by solving the task allocation MIQP~\eqref{eq:allocationalgorithm}. The aerial robots in the middle of the domain are executing the monitoring task using coverage control. Two aerial robots remain idle at their starting locations (denoted by purple circles), as all task requirements are met by the rest of the team. \ch{We introduce an environmental disturbance in the form of a region of low friction---indicated by the large blue colored area---where, as seen in Fig.~\ref{subfig:exp:2}, the motion of the ground robot allocated to the task is impeded and it cannot track the assigned target anymore.} Thanks to the capability degradation computations in Section~\ref{sec:task_perf}, this anomaly is accounted for in constraint~\eqref{eq:miqp:b} of the MIQP. In particular, Fig.~\ref{fig:cap_margins} illustrates how the capability margin corresponding to ground mobility for task 2 ($\mb{D}_{2,4}$) decreases after the failure and becomes negative. Consequently, the event triggered MIQP switches task 2 to the second configuration (showcased by $\iota_2$ in the top left corner of Fig.~\ref{subfig:exp:2}). This ensures that only aerial robots---unaffected by the slippery ground---are deployed to track target 2. Figure~\ref{subfig:exp:2} shows the two additional aerial robots joining task 2 (highlighted by the green ellipse). Furthermore, constraint~\eqref{eq:miqp:e} ensures that the stuck ground robot is not assigned to any task and it returns to it's starting location. \par  
At time $27$ seconds, a weather event affects the ability of the robots in task 2 to maintain an effective tracking quality of the target---signified by an increase in the value of $e_i$. This is depicted in Fig.~\ref{subfig:exp:3} as a foggy area surrounding target $2$ and a decreasing capability margin $\mb{D}_{2,2}$ corresponding to the ``resolution" capability in the right tile of Fig.~\ref{fig:cap_margins}. Since there are no more robots available to join task 2, the algorithm relaxes the constraints corresponding to the monitoring task, and reallocates one aerial robot to the tracking task ensuring that the overall capability margin stays above zero, while that for task 3 (depicted by $\mb{D}_{3,2}$) falls below zero. This demonstrates the ability of our algorithm to \emph{gracefully degrade performance} when necessary.\par
\begin{figure}[h]
    \centering
    \subfloat[][]{
    \includegraphics[trim={3cm 1cm 3cm 3cm},clip,width=0.249\textwidth]{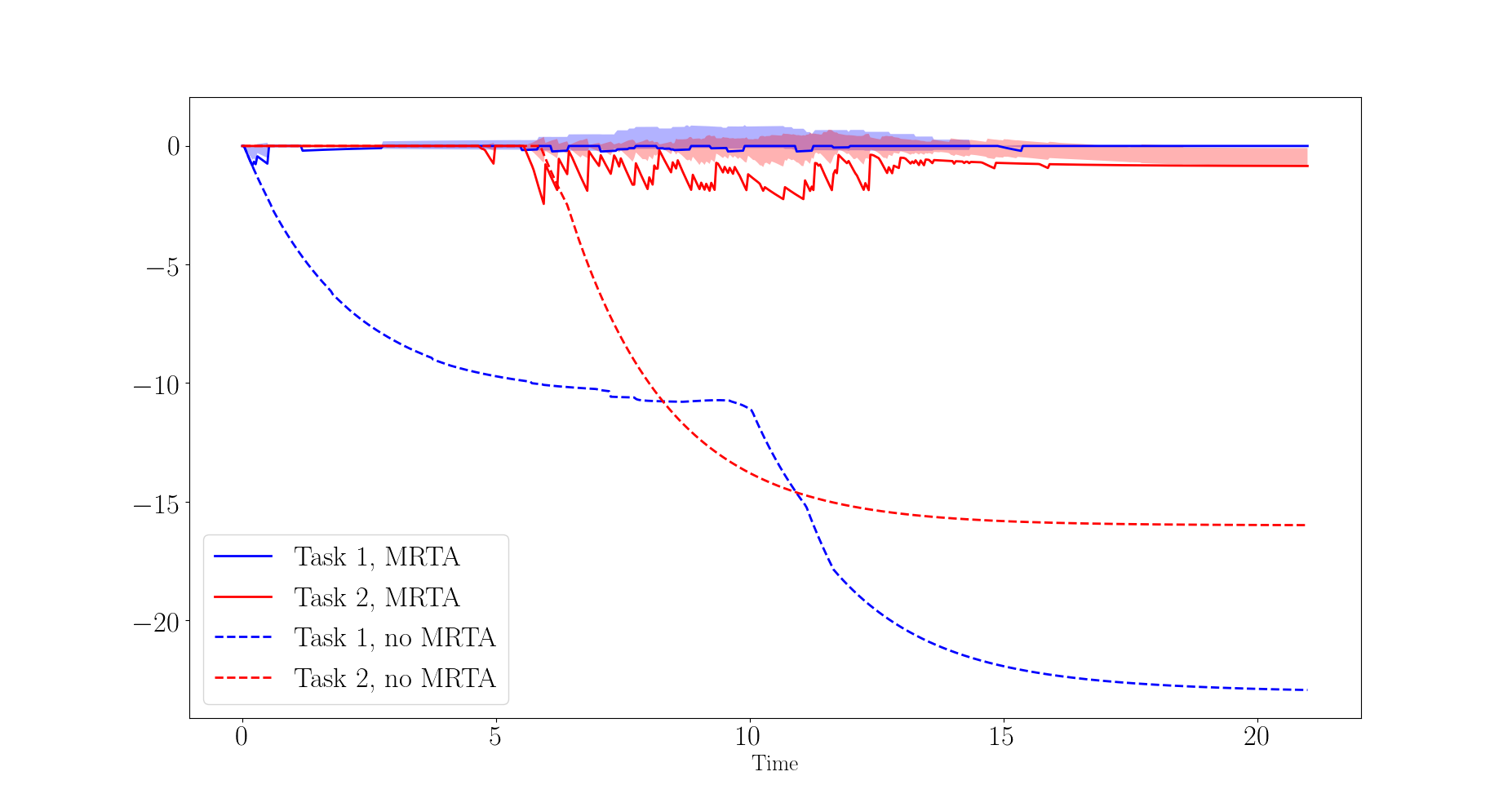}
    \label{fig:ensemble_results}}
    \subfloat[][]{
    \includegraphics[trim={3cm 1cm 3cm 3cm},clip,width=0.249\textwidth]{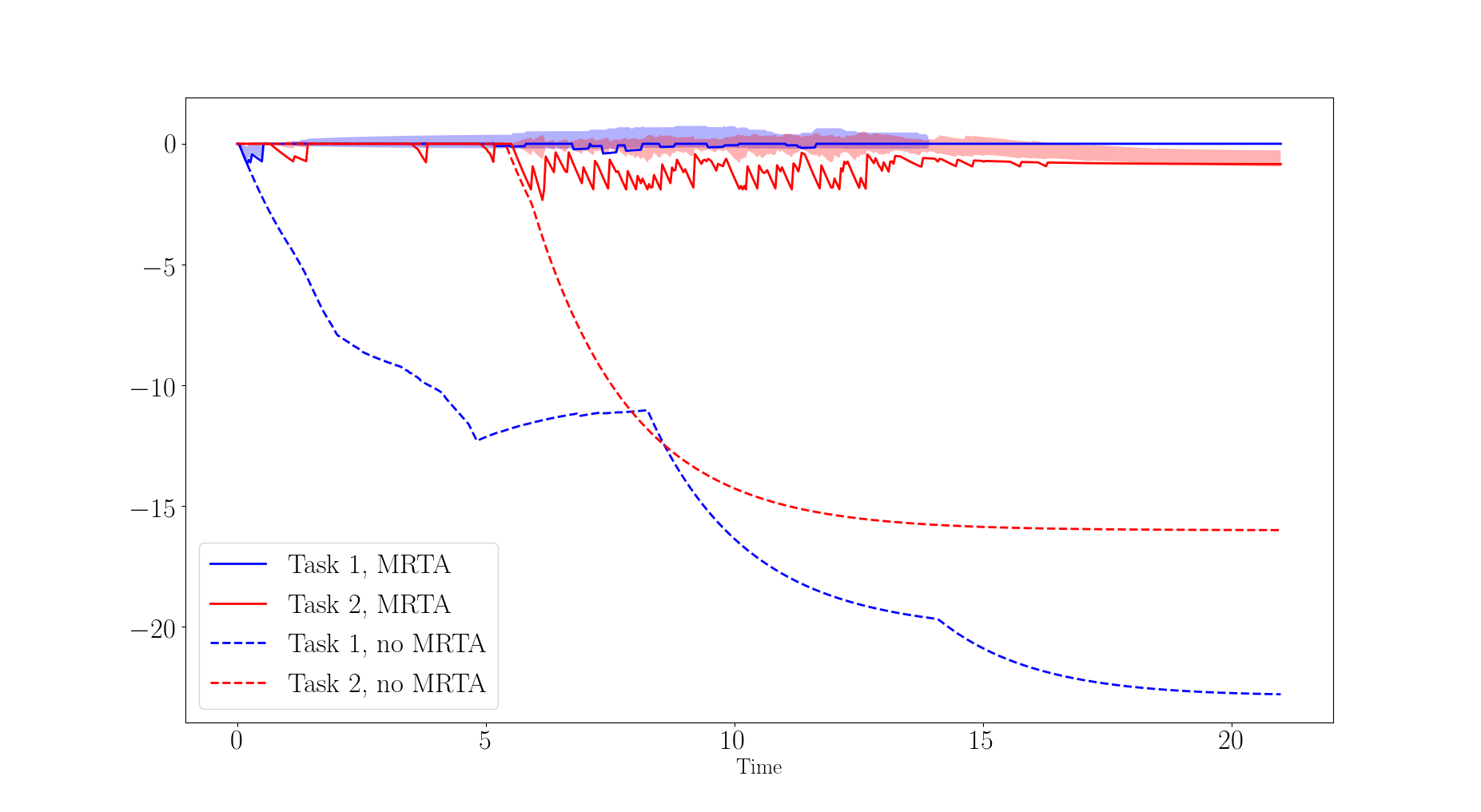}
    \label{fig:ensemble_results_param}}
    \caption{Capability margin (see Definition~\ref{defn:capab_margin}) results from 20 randomized trials of the coverage and tracking tasks with a team of 40 robots. The lines represent the lowest worst-case capability margins over all the trials for target tracking tasks 1 and 2 corresponding to two cases: with and without the event triggered task allocation algorithm. \ch{\ref{fig:ensemble_results} depicts the results from the randomization of environmental variables and \ref{fig:ensemble_results_param} considers the randomization of constants in the task allocation optimization program. In the case where resilient reallocations are regularly applied, the worst-case capability margins for the tasks remain significantly closer to 0, ensuring they are effectively executed (solid lines). The shaded regions represent the $\pm 1$ standard deviation of the results obtained over the trials. Furthermore, the results in~\ref{fig:ensemble_results_param} demonstrates that adjustments to the optimization parameters do not drastically vary the performance of the algorithm.}}
    
\end{figure}

 
In order to verify the ability of the proposed allocation algorithm to deal with large robot teams and varied environmental conditions, we ran multiple randomized trials of the coverage and target tracking mission described above with a team of 32 aerial robots and 8 ground robots (with a modified~\eqref{eqn_task_1_2_req} and~\eqref{eqn_task_3_req}). The timing corresponding to the target movement, weather events, as well as the initial positions of the robots were randomized in each of the trials. Over 20 independent trials, Fig.~\ref{fig:ensemble_results} depicts the minimum (worst case among all runs) capability margins corresponding to two cases: with and without the event-triggered resilient task allocation algorithm. For the second case, the allocation algorithm was only executed once at the beginning of each trial. As seen, the resilient allocation algorithm ensures that the capability margins remain close to or at zero, ensuring that the tasks can progress successfully, despite the environmental variations.\par
\begin{figure}[h]
    \centering
    \includegraphics[width=0.45\textwidth]{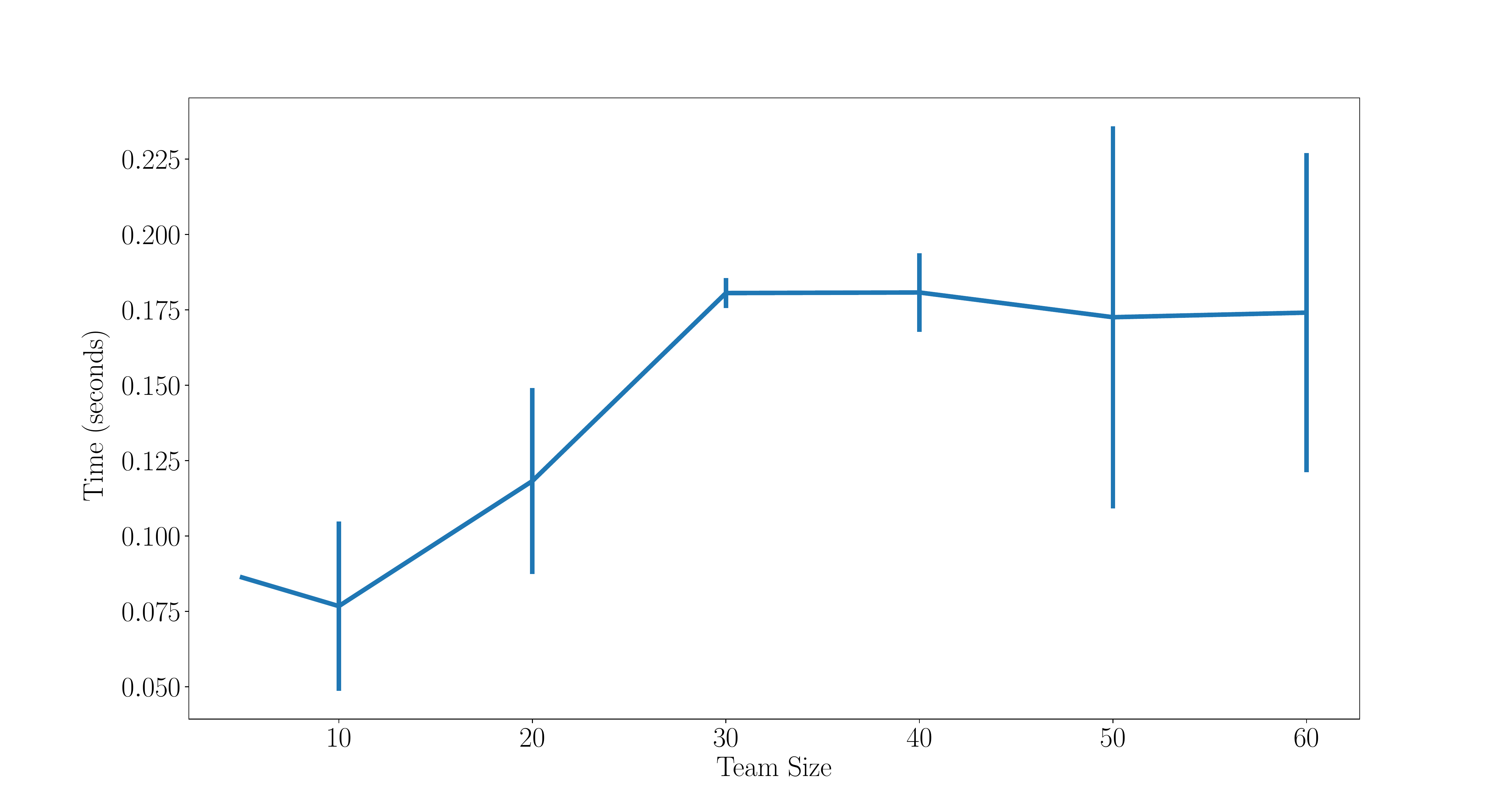}
    \caption{\ch{Computation time corresponding to the resilient task allocation MIQP for varying team sizes. As seen, the solution time remains below one second even for large teams, making the algorithm suitable for being executed in an event-triggered manner. The vertical bars represent the $\pm$ one standard deviation in the computational times over the course of the simulation.}}
    \label{fig:comp_time}
\end{figure}

 
\ch{With the aim of testing the sensitivity of the proposed task allocation paradigm with respect to the various optimization constants used in~\eqref{eq:allocationalgorithmactual}, Fig.~\ref{fig:ensemble_results_param} illustrates the results of 20 independent trials conducted over a randomized set of simulation parameters, $\mb{W_s}, \mb{w}_J, \delta_{max}$, $l$, $T$, and $\chi$. Each parameter was drawn from a normal distribution centered around their nominal value and a standard deviation equal to 40\% of the magnitude of the mean. The worst case performance as well as the variations among all the trials is depicted in Fig.~\ref{fig:ensemble_results_param}. As seen, the performance of the MIQP remains quite close among all the runs, and remains significantly better than the case when no allocations are performed. Figure~\ref{fig:comp_time} plots the average time taken to solve the MIQP for different team sizes (executed on a laptop equipped with an Intel i7-8$^{th}$ Gen CPU with 16GB of RAM). Notably, the time taken to execute the MIQP remains below one second even for large swarms---a smaller time scale compared to the rate at which new environmental disturbances might affect the system.}


\section{CONCLUSIONS} \label{sec:conc}
For heterogeneous multi-robot systems operating in dynamic conditions, we present a resilient task allocation framework which explicitly leverages information pertaining to the real-time task performance of robots when generating robot-task assignments. \ch{The task allocation framework is centralized, and relies on a communication channel between the robots and the central computer to generate the task reallocations (although not for the execution of the tasks itself). In the future, we would like to enable a distributed computation of the robot-task allocations. The proposed method also uses a simple averaging method to attribute capability degradation values within each task. Using ideas from fault detection/diagnosis, a more sophisticated mechanism can be integrated into our framework, resulting in more expressive and resilient allocations.}

\addtolength{\textheight}{-8cm}   



\bibliographystyle{unsrt}
\bibliography{references.bib}

\begin{thebibliography}{10}

\bibitem{iocchi2003distributed}
Luca Iocchi, Daniele Nardi, Maurizio Piaggio, and Antonio Sgorbissa.
\newblock Distributed coordination in heterogeneous multi-robot systems.
\newblock {\em Autonomous robots}, 15(2):155--168, 2003.

\bibitem{gunn2015dynamic}
Tyler Gunn and John Anderson.
\newblock Dynamic heterogeneous team formation for robotic urban search and
  rescue.
\newblock {\em Journal of Computer and System Sciences}, 81(3):553--567, 2015.

\bibitem{rizk2019cooperative}
Yara Rizk, Mariette Awad, and Edward~W Tunstel.
\newblock Cooperative heterogeneous multi-robot systems: a survey.
\newblock {\em ACM Computing Surveys (CSUR)}, 52(2):1--31, 2019.

\bibitem{taxonomy}
Brian~P. Gerkey and Maja~J. Matarić.
\newblock A formal analysis and taxonomy of task allocation in multi-robot
  systems.
\newblock {\em The International Journal of Robotics Research}, 23(9):939--954,
  2004.

\bibitem{taxonomy2}
G.~Ayorkor Korsah, Anthony Stentz, and M.~Bernardine Dias.
\newblock A comprehensive taxonomy for multi-robot task allocation.
\newblock {\em The International Journal of Robotics Research},
  32(12):1495--1512, 2013.

\bibitem{khamis2015multi}
Alaa Khamis, Ahmed Hussein, and Ahmed Elmogy.
\newblock Multi-robot task allocation: A review of the state-of-the-art.
\newblock In {\em Cooperative Robots and Sensor Networks 2015}, pages 31--51.
  Springer, 2015.

\bibitem{prorok2017impact}
Amanda Prorok, M~Ani Hsieh, and Vijay Kumar.
\newblock The impact of diversity on optimal control policies for heterogeneous
  robot swarms.
\newblock {\em IEEE Transactions on Robotics}, 33(2):346--358, 2017.

\bibitem{ravichandar2020strata}
Harish Ravichandar, Kenneth Shaw, and Sonia Chernova.
\newblock Strata: unified framework for task assignments in large teams of
  heterogeneous agents.
\newblock {\em Autonomous Agents and Multi-Agent Systems}, 34(2), 2020.

\bibitem{ramachandran2019resilience}
Ragesh~K Ramachandran, James~A Preiss, and Gaurav~S Sukhatme.
\newblock Resilience by reconfiguration: Exploiting heterogeneity in robot
  teams.
\newblock In {\em 2019 IEEE/RSJ International Conference on Intelligent Robots
  and Systems (IROS)}, pages 6518--6525. IEEE, 2019.

\bibitem{saulnier2017resilient}
Kelsey Saulnier, David Salda{\~n}a, Amanda Prorok, George~J Pappas, and Vijay
  Kumar.
\newblock Resilient flocking for mobile robot teams.
\newblock {\em IEEE Robotics and Automation letters}, 2(2):1039--1046, 2017.

\bibitem{emam2020adaptive}
Yousef Emam, Siddharth Mayya, Gennaro Notomista, Addison Bohannon, and Magnus
  Egerstedt.
\newblock Adaptive task allocation for heterogeneous multi-robot teams with
  evolving and unknown robot capabilities.
\newblock {\em arXiv preprint arXiv:2003.03344}, 2020.

\bibitem{notomista2019optimal}
Gennaro Notomista, Siddharth Mayya, Seth Hutchinson, and Magnus Egerstedt.
\newblock An optimal task allocation strategy for heterogeneous multi-robot
  systems.
\newblock In {\em 2019 18th European Control Conference (ECC)}, pages
  2071--2076. IEEE, 2019.

\bibitem{parker1994heterogeneous}
Lynne~E Parker.
\newblock Heterogeneous multi-robot cooperation.
\newblock Technical report, Massachusetts Inst of Tech Cambridge Artificial
  Intelligence Lab, 1994.

\bibitem{marcolino2013multi}
Leandro~Soriano Marcolino, Albert~Xin Jiang, and Milind Tambe.
\newblock Multi-agent team formation: diversity beats strength?
\newblock In {\em IJCAI}, volume~13. Citeseer, 2013.

\bibitem{vig2006multi}
Lovekesh Vig and Julie~A Adams.
\newblock Multi-robot coalition formation.
\newblock {\em IEEE transactions on robotics}, 22(4):637--649, 2006.

\bibitem{cortes2004coverage}
Jorge Cortes, Sonia Martinez, Timur Karatas, and Francesco Bullo.
\newblock Coverage control for mobile sensing networks.
\newblock {\em IEEE Transactions on robotics and Automation}, 20(2):243--255,
  2004.

\bibitem{pimenta2009simultaneous}
Luciano~CA Pimenta, Mac Schwager, Quentin Lindsey, Vijay Kumar, Daniela Rus,
  Renato~C Mesquita, and Guilherme~AS Pereira.
\newblock Simultaneous coverage and tracking (scat) of moving targets with
  robot networks.
\newblock In {\em Algorithmic foundation of robotics VIII}, pages 85--99.
  Springer, 2009.

\bibitem{oh2015survey}
Kwang-Kyo Oh, Myoung-Chul Park, and Hyo-Sung Ahn.
\newblock A survey of multi-agent formation control.
\newblock {\em Automatica}, 53:424--440, 2015.

\bibitem{prorok2016formalizing}
Amanda Prorok, M~Ani Hsieh, and Vijay Kumar.
\newblock Formalizing the impact of diversity on performance in a heterogeneous
  swarm of robots.
\newblock In {\em 2016 IEEE International Conference on Robotics and Automation
  (ICRA)}, pages 5364--5371. IEEE, 2016.

\bibitem{turpin2014capt}
Matthew Turpin, Nathan Michael, and Vijay Kumar.
\newblock Capt: Concurrent assignment and planning of trajectories for multiple
  robots.
\newblock {\em The International Journal of Robotics Research}, 33(1):98--112,
  2014.

\bibitem{cortes2017coordinated}
Jorge Cort{\'e}s and Magnus Egerstedt.
\newblock Coordinated control of multi-robot systems: A survey.
\newblock {\em SICE Journal of Control, Measurement, and System Integration},
  10(6):495--503, 2017.

\bibitem{burgard2000collaborative}
Wolfram Burgard, Mark Moors, Dieter Fox, Reid Simmons, and Sebastian Thrun.
\newblock Collaborative multi-robot exploration.
\newblock In {\em Proceedings 2000 ICRA. Millennium Conference. IEEE
  International Conference on Robotics and Automation. Symposia Proceedings
  (Cat. No. 00CH37065)}, volume~1, pages 476--481. IEEE, 2000.

\bibitem{yan2013survey}
Zhi Yan, Nicolas Jouandeau, and Arab~Ali Cherif.
\newblock A survey and analysis of multi-robot coordination.
\newblock {\em International Journal of Advanced Robotic Systems}, 10(12):399,
  2013.

\bibitem{shishika2019team}
Daigo Shishika, James Paulos, Michael~R Dorothy, M~Ani Hsieh, and Vijay Kumar.
\newblock Team composition for perimeter defense with patrollers and defenders.
\newblock In {\em 2019 IEEE 58th Conference on Decision and Control (CDC)},
  pages 7325--7332. IEEE, 2019.

\bibitem{del2017mixed}
Alberto Del~Pia, Santanu~S Dey, and Marco Molinaro.
\newblock Mixed-integer quadratic programming is in np.
\newblock {\em Mathematical Programming}, 162(1-2):225--240, 2017.

\bibitem{coppeliaSim}
E.~Rohmer, S.~P.~N. Singh, and M.~Freese.
\newblock Coppeliasim (formerly v-rep): a versatile and scalable robot
  simulation framework.
\newblock In {\em Proc. of The International Conference on Intelligent Robots
  and Systems (IROS)}, 2013.
\newblock www.coppeliarobotics.com.

\bibitem{coumans2013bullet}
Erwin Coumans et~al.
\newblock Bullet physics library.
\newblock {\em Open source: bulletphysics. org}, 15(49):5, 2013.

\end{thebibliography}
\end{document}